\documentclass[10pt,conference]{IEEEtran}
\IEEEoverridecommandlockouts
\usepackage{cite}
\usepackage{float}
\usepackage{amsthm,amsmath,amssymb,amsfonts}
\usepackage{mathtools}
\usepackage{algorithmic}
\usepackage{graphicx}
\usepackage{subcaption}
\usepackage{textcomp}
\usepackage{xcolor}
\usepackage{tabularx,ragged2e,booktabs}
\usepackage{multirow}
\usepackage{rotating}
\usepackage[utf8]{inputenc}
\usepackage{dblfloatfix}
\usepackage{ulem}
\usepackage[hang,flushmargin]{footmisc}
\usepackage[font=small,labelfont=bf]{caption}
\usepackage{seqsplit}
\usepackage[linesnumbered,ruled,vlined]{algorithm2e}
\usepackage{url}
\usepackage{xurl}

\def\BibTeX{{\rm B\kern-.05em{\sc i\kern-.025em b}\kern-.08em
    T\kern-.1667em\lower.7ex\hbox{E}\kern-.125emX}}

\begin{document}

\title{Detection and Resolution of Periodic Artifacts in OpenDP's Discrete Laplace Sampler}

\author{
    \IEEEauthorblockN{Cesare Gerolimetto Fabrello\IEEEauthorrefmark{1}\IEEEauthorrefmark{2}, Valeria Rossi\IEEEauthorrefmark{1}\IEEEauthorrefmark{2}, Alberto Trombetta\IEEEauthorrefmark{1}, and Massimo Caccia\IEEEauthorrefmark{1}\IEEEauthorrefmark{2}}
    \IEEEauthorblockA{\IEEEauthorrefmark{1}Università degli Studi dell'Insubria}
    \IEEEauthorblockA{\IEEEauthorrefmark{2}Random Power Srl}
}

\maketitle
    
\begin{abstract}
Differential privacy implementations rely on precise sampling from noise distributions to provide formal privacy guarantees. We report the discovery of systematic artifacts in OpenDP's discrete Laplace sampler that manifest as periodic distortions in the output distribution. Through systematic testing, we trace these artifacts to a faulty implementation in the rational arithmetic library used by the \texttt{bernoulli\_exp1} function, a low-level primitive that implements sampling from $\text{Bernoulli}(\exp(-x))$ distributions. We present a diagnostic methodology that isolates the faulty component in the nested sampling hierarchy and propose an alternative implementation based on exact rational arithmetic that eliminates the artifacts. Statistical validation with $10^6$ samples confirms that the corrected sampler produces outputs indistinguishable from the theoretical distribution at the tested precision level.
\end{abstract}


\thispagestyle{plain}
\pagestyle{plain}
\pagenumbering{gobble}

\section{Introduction}
\label{sec:intro}

Differential privacy has emerged as the gold standard for privacy-preserving data analysis \cite{dp_book}. Central to any differential privacy implementation is the noise addition mechanism, which relies on sampling from carefully calibrated probability distributions. The correctness of these sampling procedures is critical. Even small deviations can compromise privacy guarantees or degrade utility. While the mathematical foundations are well understood, their realization on finite-precision hardware introduces subtle challenges. For example, Mironov \cite{mironov_lsb} demonstrated that floating-point arithmetic can leak information through the least significant bits. Our work extends this to exact sampling implementations, where numerical precision issues manifest as systematic artifacts.

OpenDP \cite{opendp_website} is a widely used open-source library implementing the discrete Laplace mechanism following Canonne, Kamath, and Steinke (CKS20) \cite{discrete_gauss_opendp}. This construction provides provably correct samplers through nested primitives with formal correctness guarantees. However, implementation details can introduce unexpected behaviors.

In this work, we report systematic artifacts observed in OpenDP's discrete Laplace sampler while empirically investigating the role of randomness in Differential Privacy. These artifacts manifest as periodic distortions in the empirical distribution that persist across different privacy parameters and entropy sources, suggesting a fundamental implementation issue rather than statistical fluctuation or input randomness quality.

\begin{figure}[b]
\centering
\includegraphics[width=0.9\columnwidth]{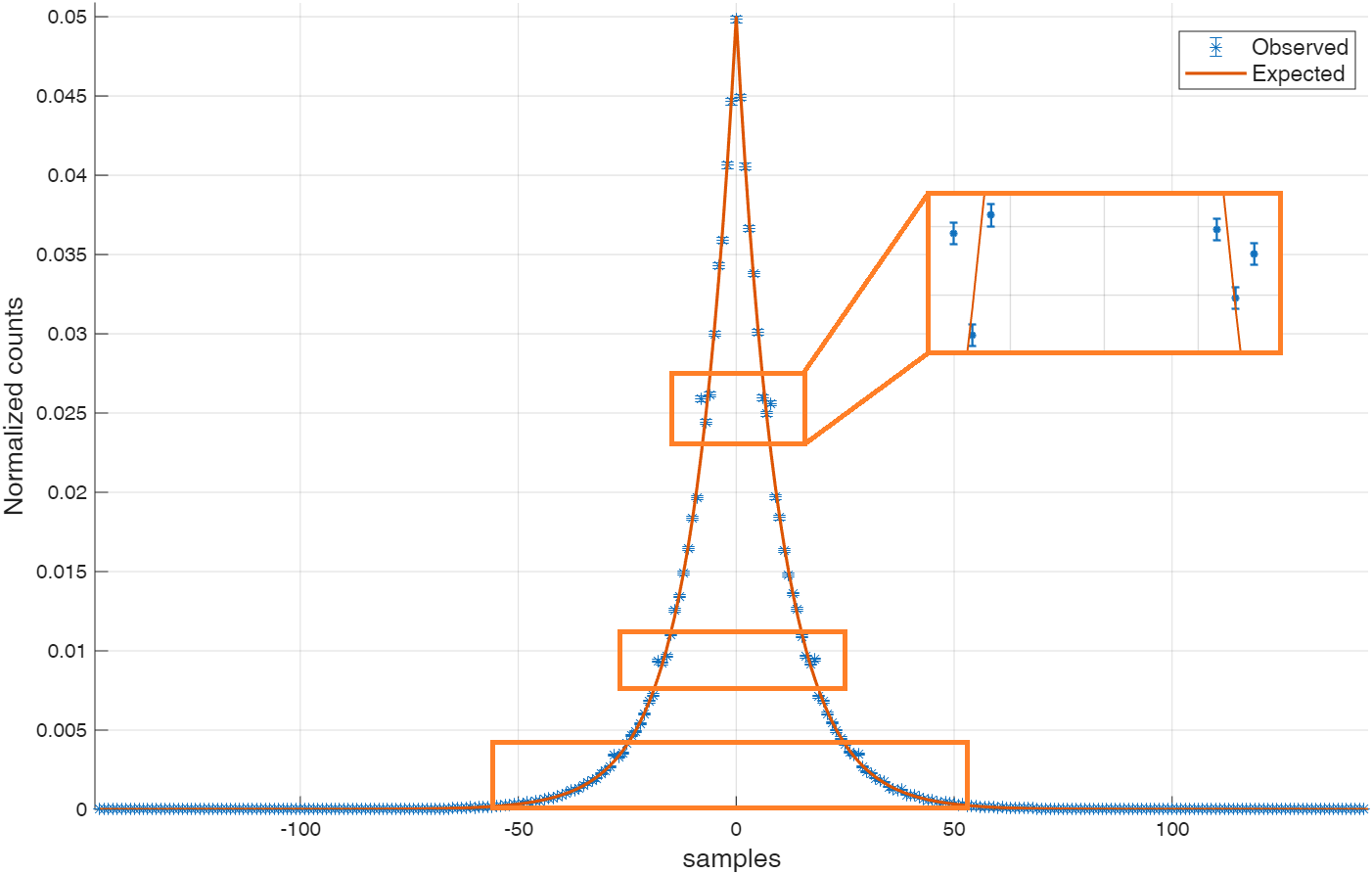}
\caption{Discrete Laplace distribution generated by OpenDP's sampler showing the characteristic periodic artifacts. The insets on the plot show a zoomed region around the artifacts, with error bars representing $1\sigma$ confidence intervals.}
\label{fig:discrete_laplace_bad}
\end{figure}

\underline{\textbf{Our contributions:}} We present a systematic methodology for diagnosing sampling artifacts in hierarchical implementations, successfully isolating the root cause to numerical precision issues in a low-level Bernoulli sampler. We propose an alternative implementation based on exact rational arithmetic that eliminates the artifacts while maintaining the theoretical guarantees of the original algorithm. Our approach demonstrates the importance of rigorous statistical testing in validating differential privacy implementations.

\section{OpenDP Library}
\label{sec:opendp}

OpenDP \cite{opendp} is an open-source differential privacy library providing implementations of standard mechanisms. The discrete 
Laplace sampler follows the construction of CKS20 \cite{discrete_gauss_opendp}, decomposing the sampling  procedure into a hierarchy of primitives with formal correctness proofs. This modular approach enables verification of individual components while providing a clear separation between mathematical guarantees and implementation details.

\begin{figure*}[t]
\centering
\begin{subfigure}[b]{0.4\textwidth}
\centering
\includegraphics[width=\textwidth]{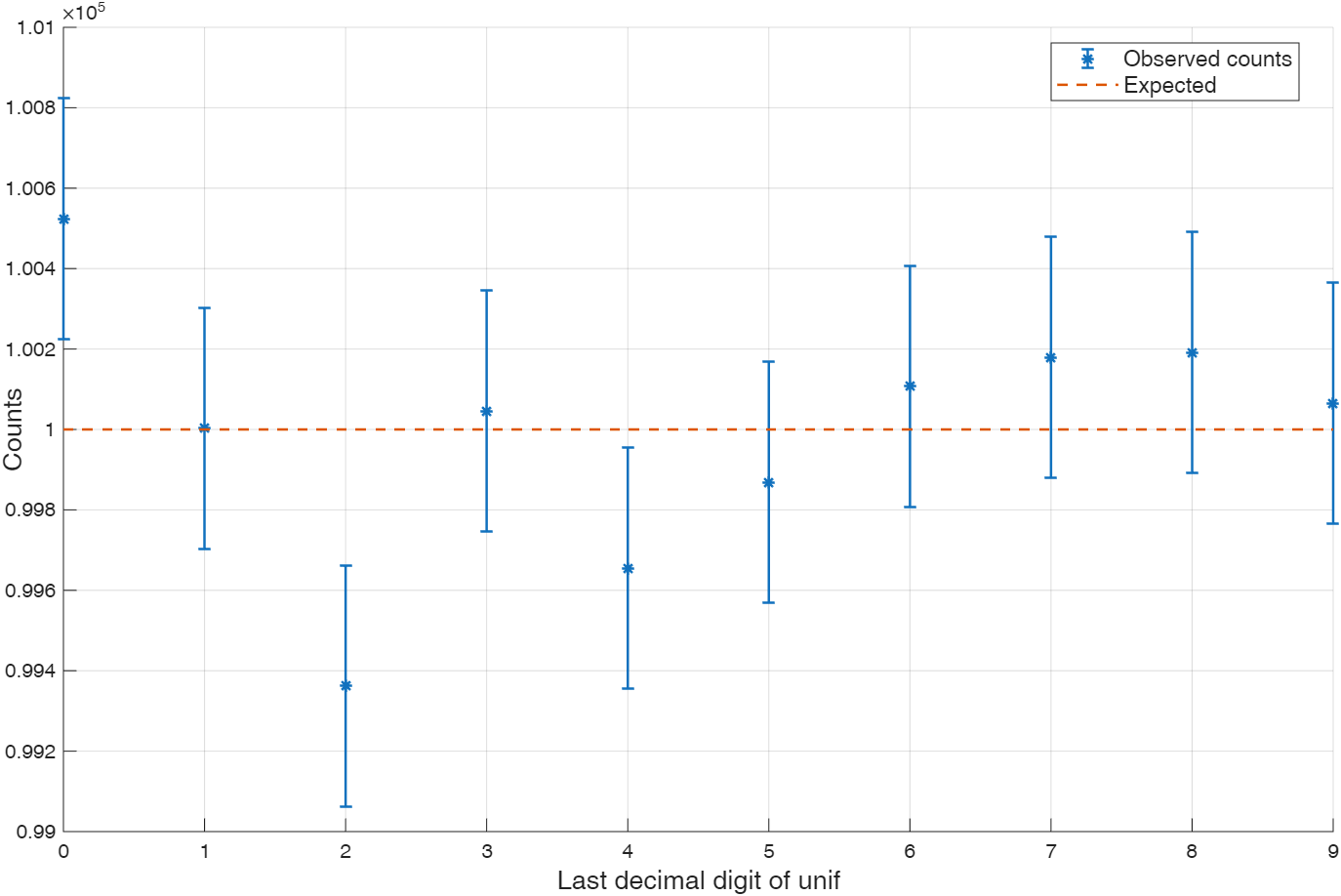}
\caption{Uniformity fit on the last digit of sampled values.}
\label{fig:uniformity_last_digit_bad}
\end{subfigure}
\hfill
\begin{subfigure}[b]{0.4\textwidth}
\centering
\includegraphics[width=\textwidth]{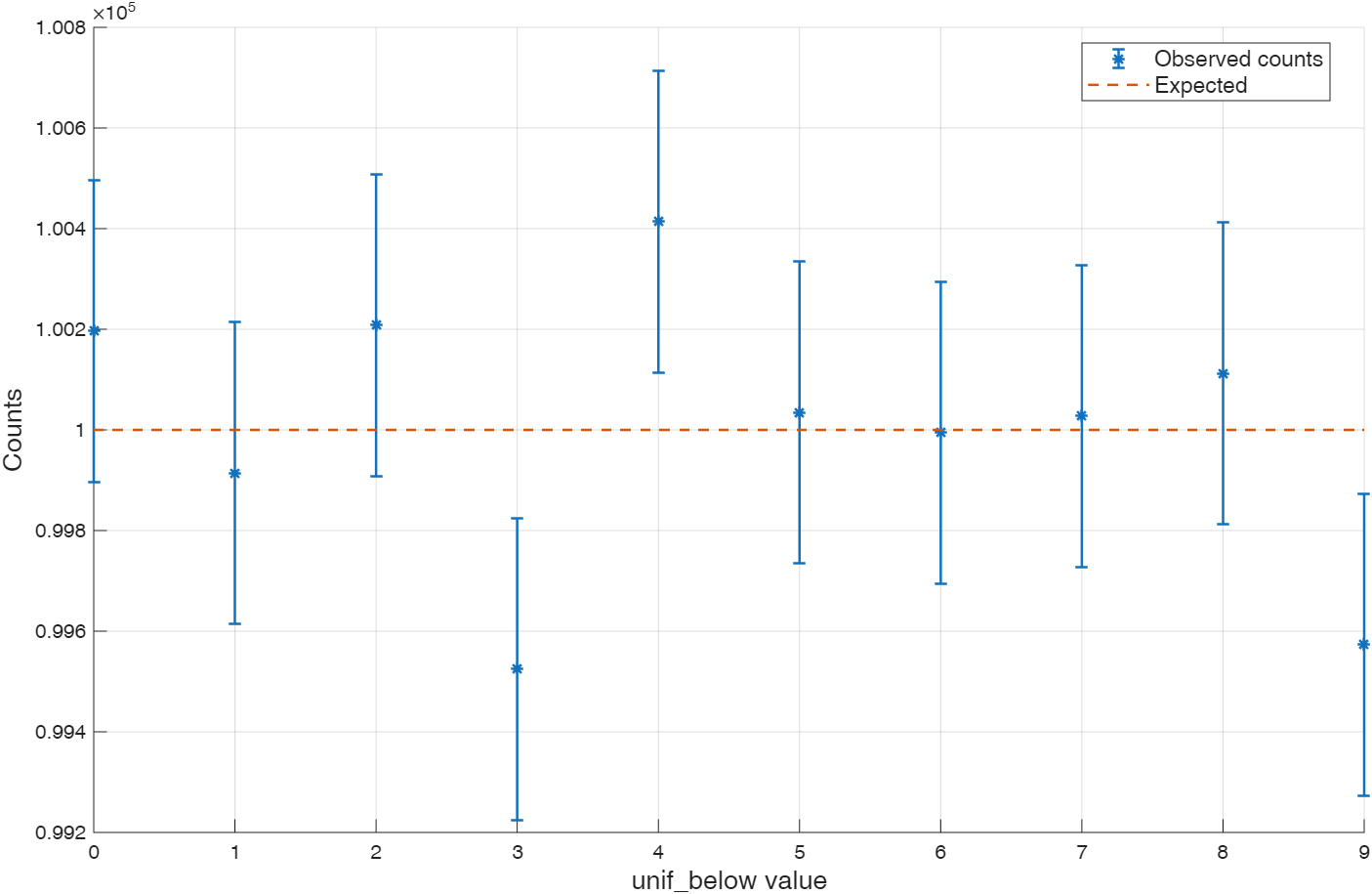}
\caption{\texttt{sample\_uniform\_ubig\_below} uniformity fit with upper bound 10.}
\label{fig:unif_below_bad}
\end{subfigure}
\caption{Uniformity tests on base sampling primitives. Left: last digit distribution of sampled uniform values ($\chi^2 = 9.04$, dof = 9, $p = 0.433$). Right: \texttt{sample\_uniform\_ubig\_below} with upper bound 10 ($\chi^2 = 6.83$, dof = 9, $p = 0.654$). Both tests show $\chi^2_{\text{red}} \approx 1$ and $p > 0.4$, confirming the entropy source operates correctly.}
\label{fig:uniformity_bad}

\begin{subfigure}[b]{0.4\textwidth}
\centering
\includegraphics[width=\textwidth]{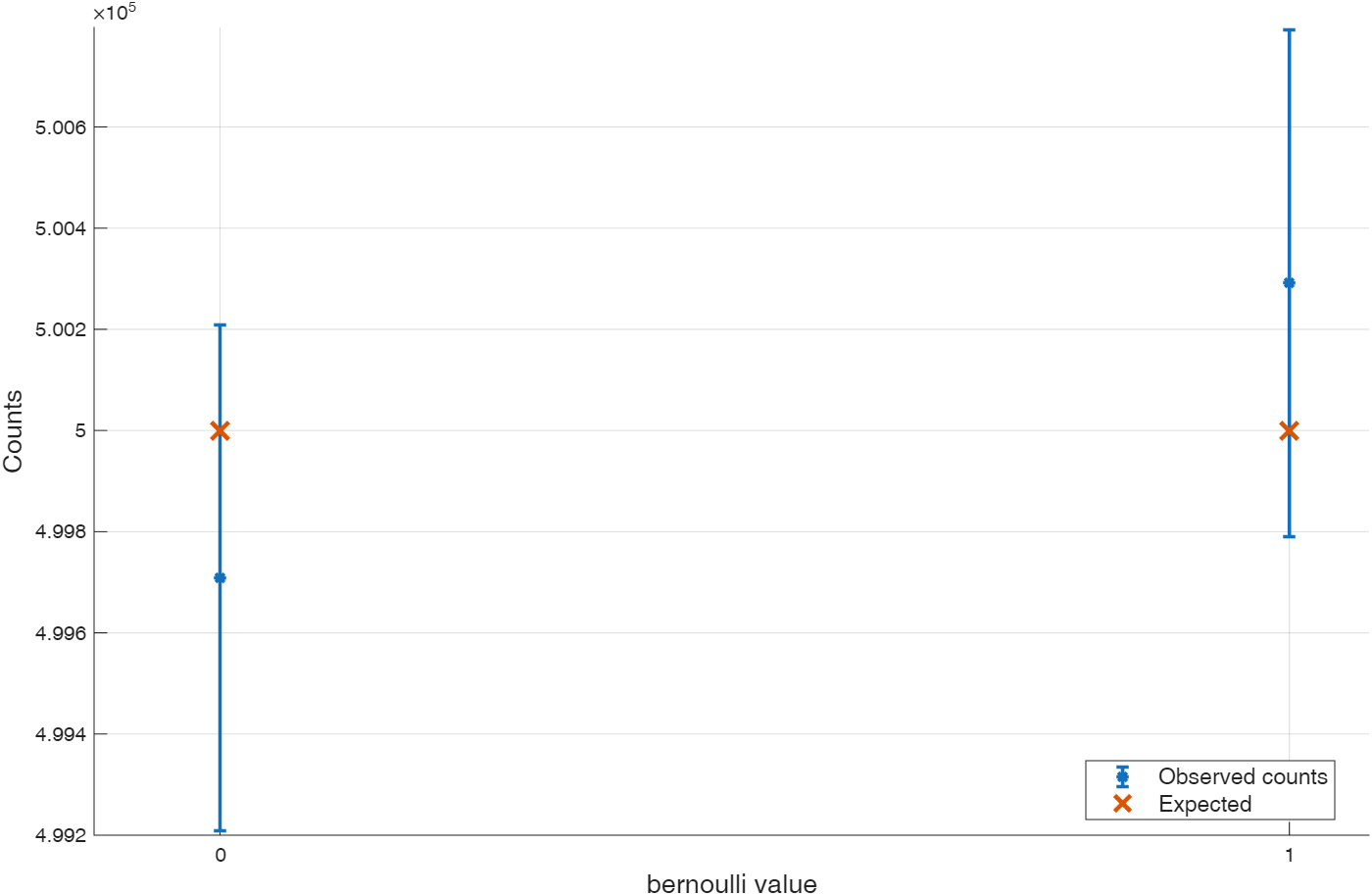}
\caption{Standard Bernoulli: $z = 0.58$, $p = 0.56$.}
\label{fig:bernoulli_standard_bad}
\end{subfigure}
\hfill
\begin{subfigure}[b]{0.4\textwidth}
\centering
\includegraphics[width=\textwidth]{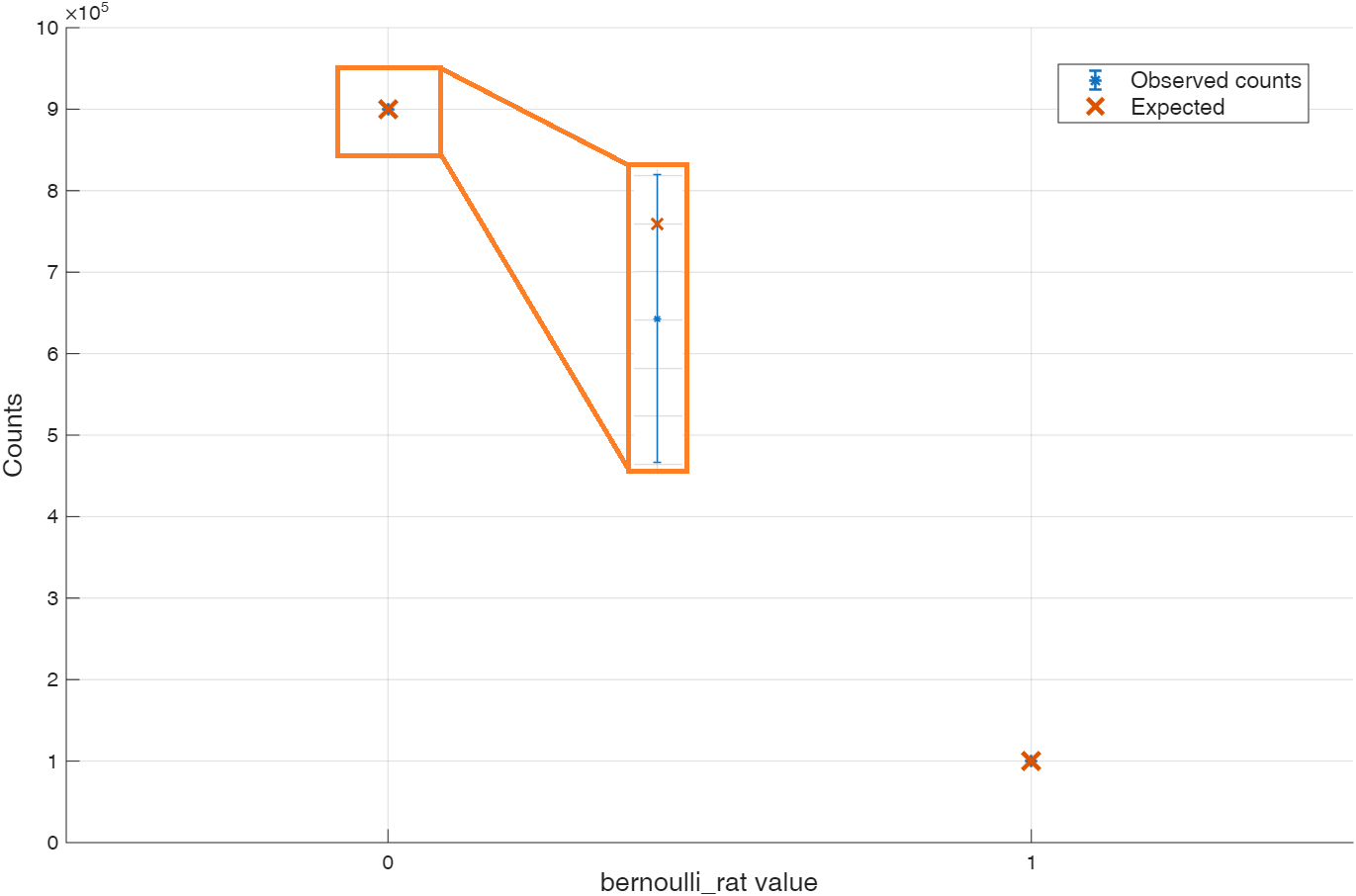}
\caption{Rational Bernoulli: $z = 0.66$, $p = 0.51$.}
\label{fig:bernoulli_rational_bad}
\end{subfigure}
\caption{Bernoulli samplers that draw directly from the entropy source. Left: standard Bernoulli with $p = 0.5$ ($\hat{p} = 0.5003$, $p_{\text{th}} = 0.5000$, $\chi^2 = 0.339$, $p = 0.561$). Right: rational Bernoulli with $p = 0.1$ ($\hat{p} = 0.1002$, $p_{\text{th}} = 0.1000$, $\chi^2 = 0.431$, $p = 0.511$). Both pass goodness-of-fit tests with $\chi^2_{\text{red}} < 1$ and $p > 0.5$. Insets show  $1\sigma$ confidence intervals.}
\label{fig:bernoulli_bad}

\vspace{-0.5cm}
\end{figure*}

\section{Artifact Discovery}
\label{sec:artifacts}

\subsection{Experimental Setup}

All experiments were conducted on a high-performance server with dual Intel Xeon Gold 5416S processors, providing 64 logical cores (32 physical cores with hyper-threading) operating between 800 MHz and 4.0 GHz. The system has 128 GB of RAM organized in a NUMA architecture with two nodes for efficient memory allocation.

These resources were required by our experimental demands: millions of iterations for each combination of parameters and randomness sources. Samples were gathered using OpenDP v.0.14.0, with all post-processing and statistical analysis performed in MATLAB. Codebase is available on a dedicated Github repository\footnote{https://github.com/grlcsr/dp\_analysis}.

\subsection{Empirical Distribution Reconstruction}

We collected $N = 10^6$ samples from OpenDP's discrete Laplace sampler to reconstruct empirical distributions. Our query counts the number of individuals in a dataset, which has sensitivity $\Delta s = 1$. Throughout this analysis, we focus on $\varepsilon = 0.1$, which yields a Laplace scale of $b = 1/\varepsilon = 10$, but the results are reproducible with different parameter choices.

Figure \ref{fig:discrete_laplace_bad} shows the reconstructed histogram using a bin size of 1. The samples exhibit evenly spaced artifacts that suggest a systematic implementation issue rather than statistical fluctuation. The choice of unitary bin size is critical: larger bins smooth over these artifacts and obscure the underlying problem. The large sample size also contributes to highlight the discrepancies.

These artifacts exhibit clear symmetry with respect to the distribution center and a periodic pattern whose spacing depends on the Laplace scale. Crucially, they appear regardless of which entropy source feeds the library, including high-quality quantum random number generators and OpenDP's native source. This strongly suggests the issue originates in the internal sampling logic rather than in the external randomness.

\begin{figure*}[t]
\centering
\begin{subfigure}[b]{0.4\textwidth}
\centering
\includegraphics[width=\textwidth]{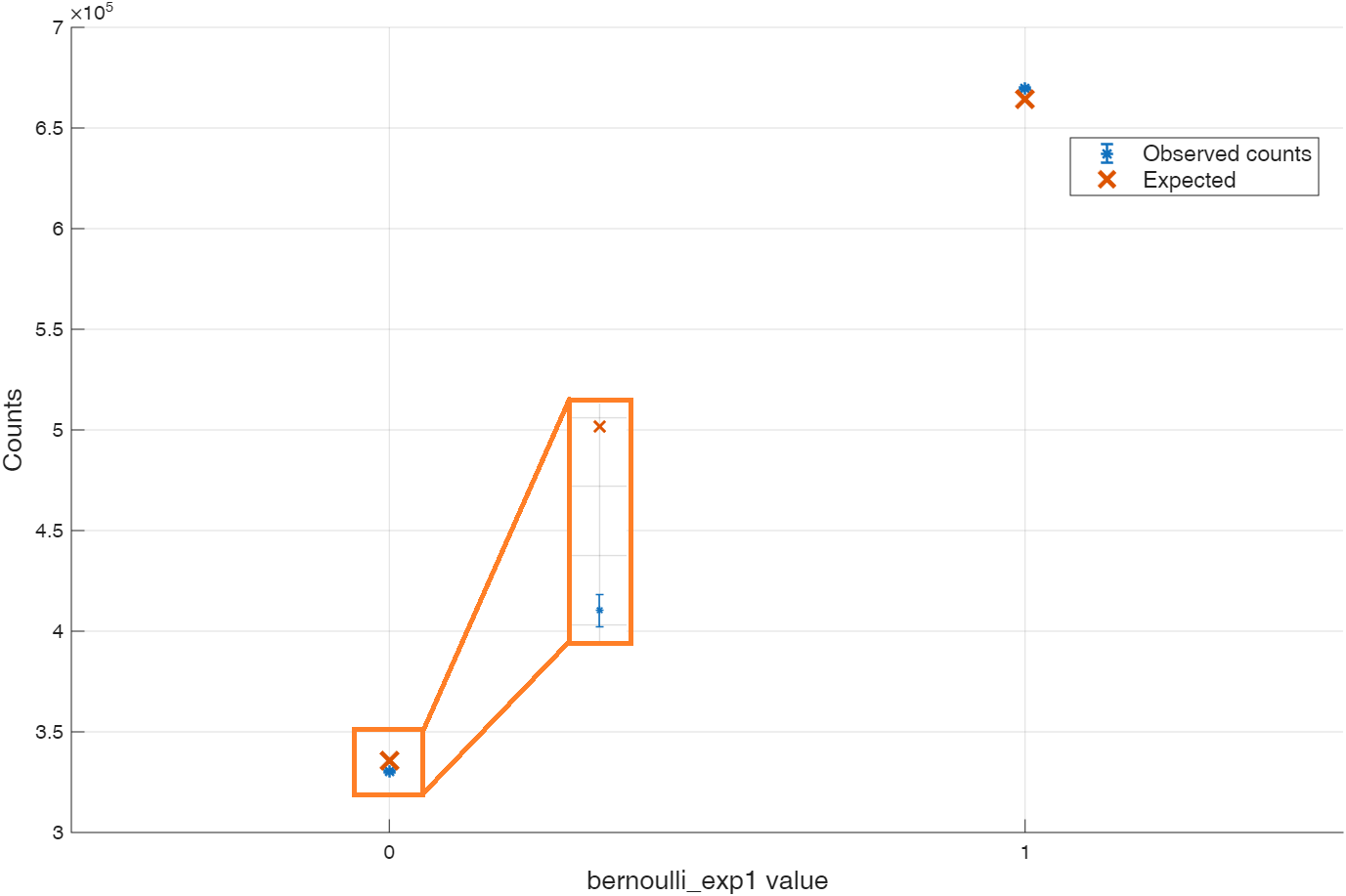}
\caption{\texttt{bernoulli\_exp1}: $z = 11.3$, $p \approx 10^{-29}$.}
\label{fig:bernoulli_exp1_bad}
\end{subfigure}
\hfill
\begin{subfigure}[b]{0.4\textwidth}
\centering
\includegraphics[width=\textwidth]{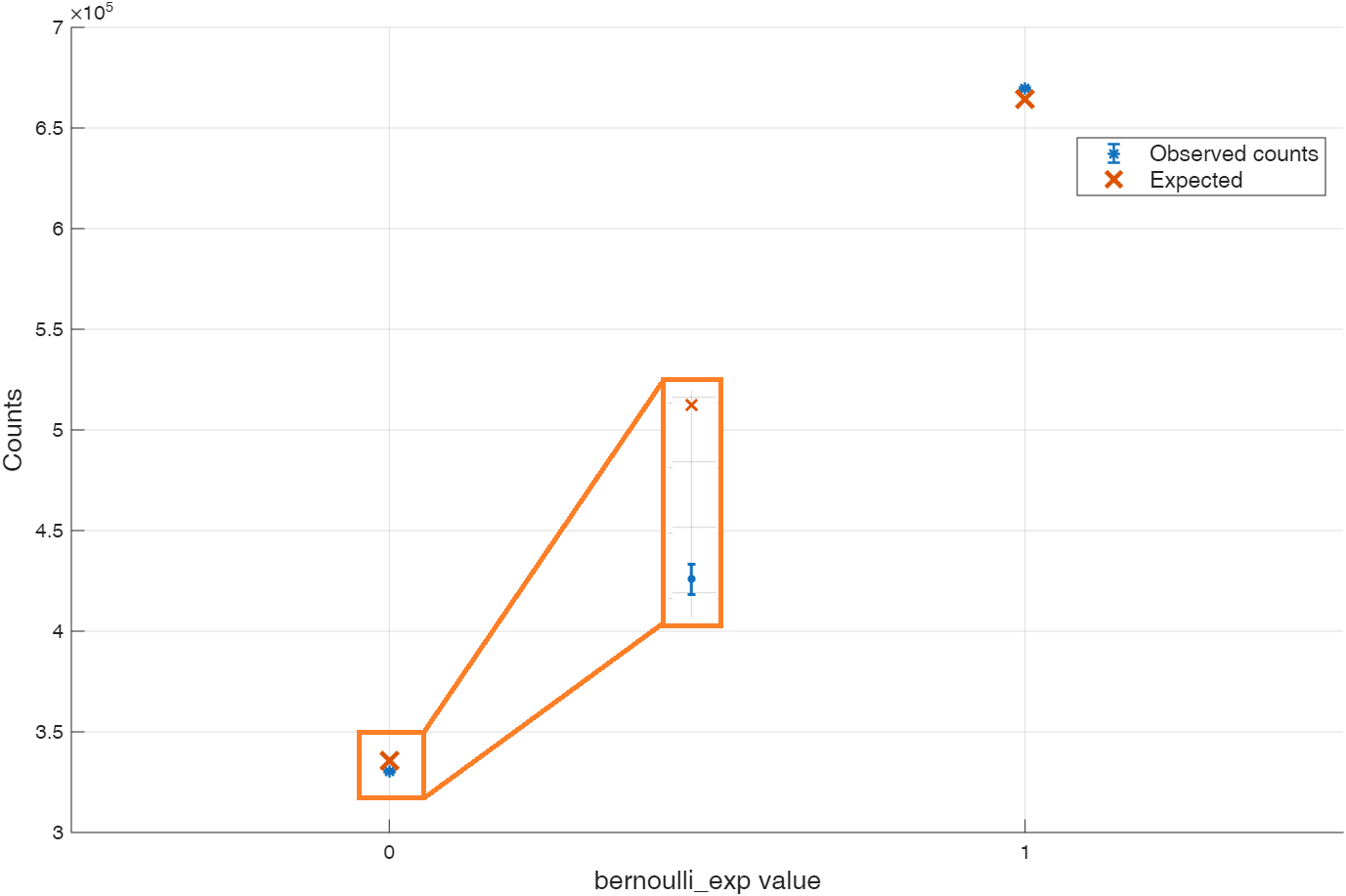}
\caption{\texttt{bernoulli\_exp}: $z = 11.3$, $p \approx 10^{-29}$.}
\label{fig:bernoulli_exp_bad}
\end{subfigure}
\caption{Exponential Bernoulli samplers showing significant deviations from the theoretical model. Left: \texttt{bernoulli\_exp} ($\hat{p} = 0.6696$, $p_{\text{th}} = 0.6643$, $\chi^2 = 128$, $p = 0$). Right: \texttt{bernoulli\_exp1} ($\hat{p} = 0.6696$, $p_{\text{th}} = 0.6643$, $\chi^2 = 128$, $p = 0$). Both fail goodness-of-fit tests with $\chi^2 \approx 128$ and $p \approx 0$. Insets zoom on the $1\sigma$ confidence intervals.}
\label{fig:bernoulli_exps_bad}

\begin{subfigure}[b]{0.4\textwidth}
\centering
\includegraphics[width=\textwidth]{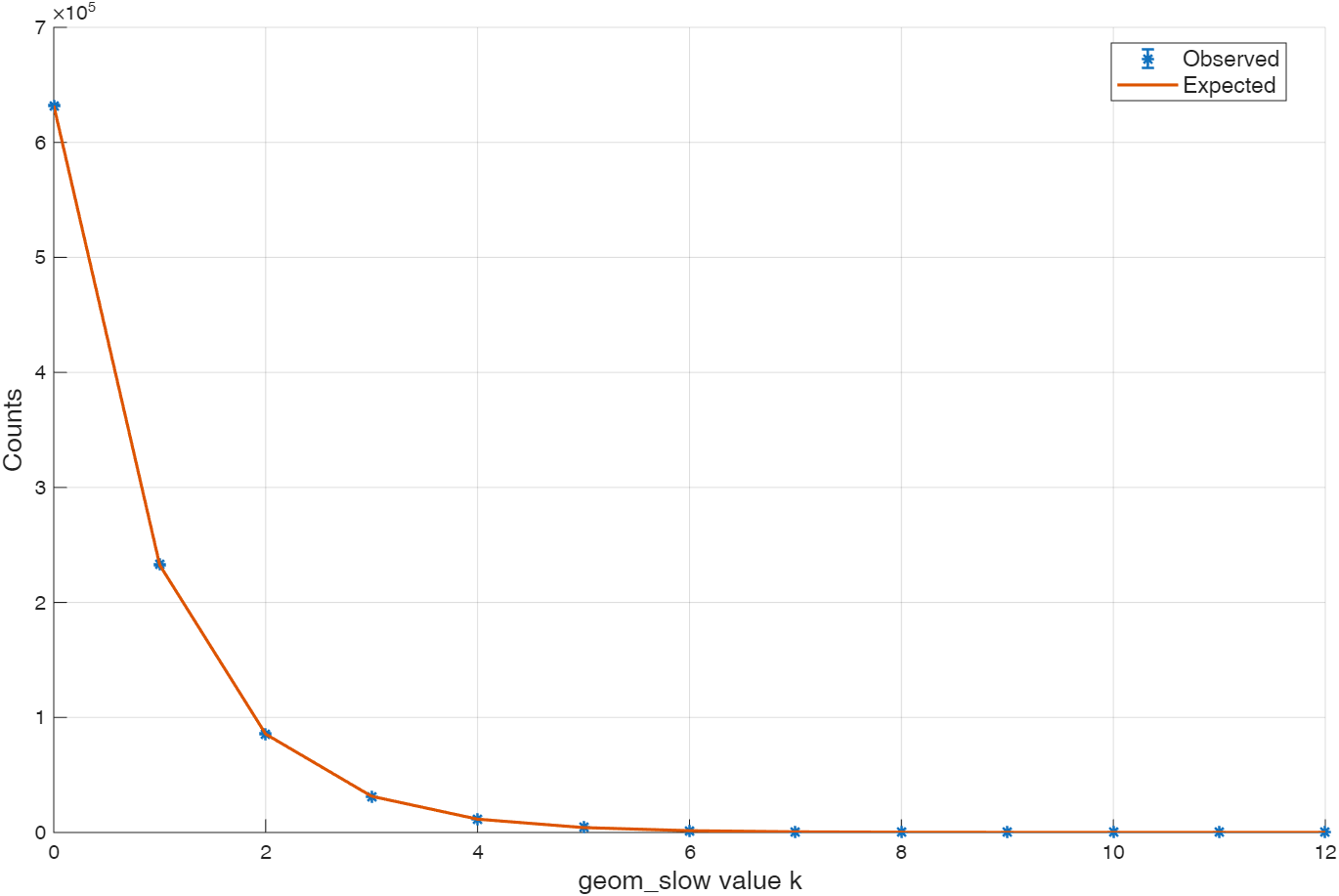}
\caption{Geometric slow: good fit to the theoretical model.}
\label{fig:geometric_slow_bad}
\end{subfigure}
\hfill
\begin{subfigure}[b]{0.4\textwidth}
\centering
\includegraphics[width=\textwidth]{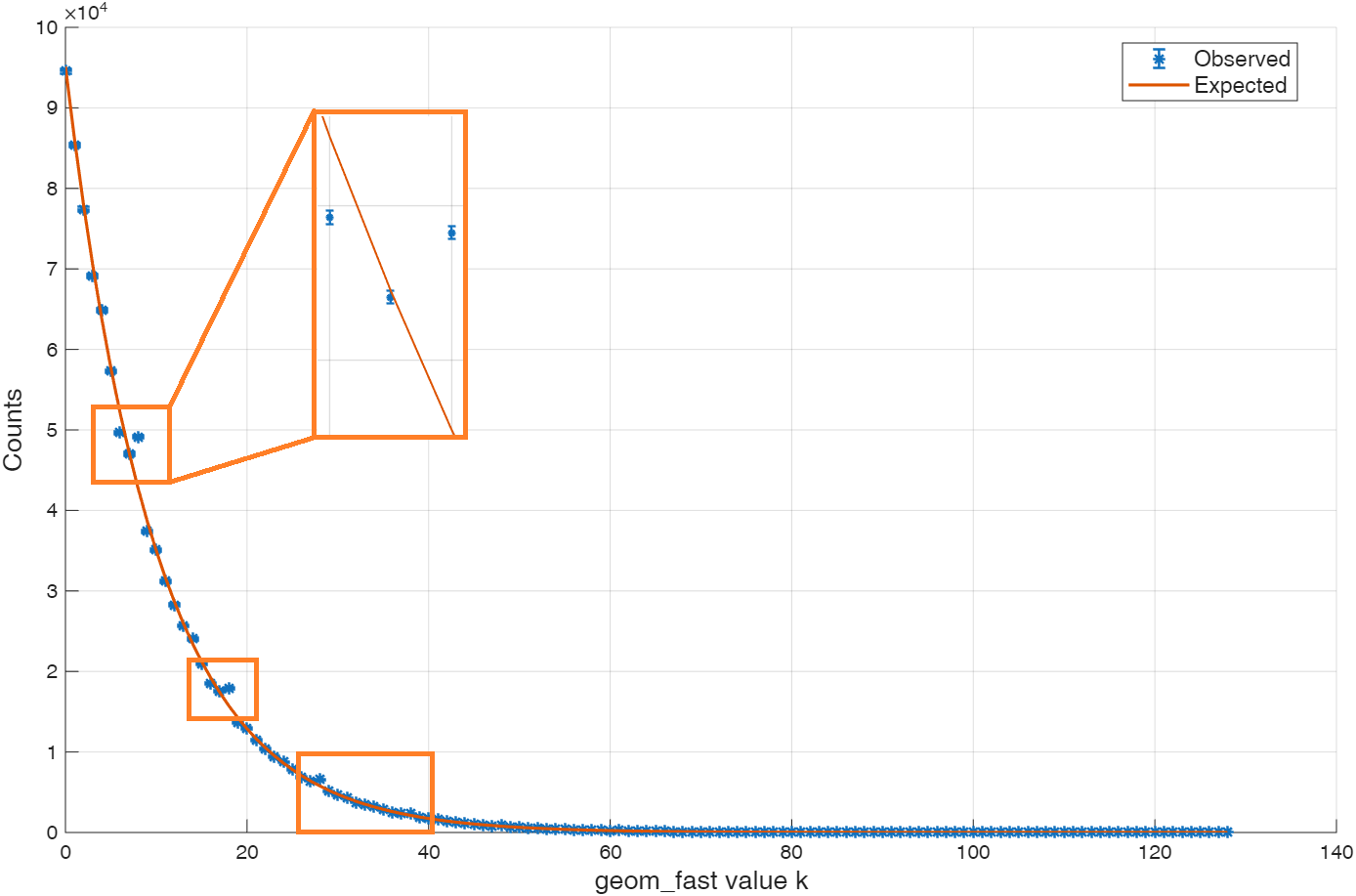}
\caption{Geometric fast: fails with visible periodic artifacts.}
\label{fig:geometric_fast_bad}
\end{subfigure}
\caption{Geometric sampler distributions. Left: slow implementation (mean $z = 0.96$, $p = 0.34$; $\chi^2 = 15$, $p = 0.26$) passes with good fit to the theoretical model. Right: fast implementation (mean $z = 3.99$, $p = 6.6 \times 10^{-5}$; $\chi^2 = 1.9 \times 10^3$, $p = 0$) fails with visible periodic artifacts similar to those observed in the final Laplace distribution. Insets zoom on the $1\sigma$ confidence intervals.}
\label{fig:geometric_bad}

\vspace{-0.5cm}
\end{figure*}

\section{Implementation Structure}
\label{sec:implementation}

According to OpenDP's documentation \cite{opendp}, the implementation follows the construction described by Canonne, Kamath, and Steinke \cite{discrete_gauss_opendp}. The sampling logic is decomposed into nested primitives that call each other in a tree-like hierarchy:

\begin{itemize}
\item \texttt{\seqsplit{sample\_discrete\_laplace}}: top-level sampler for $\text{Lap}(0,\texttt{scale})$, which draws a random sign via \texttt{\seqsplit{sample\_standard\_bernoulli}} and a magnitude via \texttt{sample\_geometric\_exp\_fast}

\item \texttt{\seqsplit{sample\_geometric\_exp\_fast}}: geometric sampler with parameter $\exp(-x)$, which repeatedly calls \texttt{\seqsplit{sample\_uniform\_ubig\_below}} and \texttt{\seqsplit{sample\_bernoulli\_exp}}

\item \texttt{\seqsplit{sample\_geometric\_exp\_slow}}: reference geometric sampler defined as a sequence of independent $\text{Bernoulli}(\exp(-x))$ trials

\item \texttt{\seqsplit{sample\_standard\_bernoulli}}: $\text{Bernoulli}(0.5)$ implemented by extracting a single random bit

\item \texttt{\seqsplit{sample\_bernoulli\_exp}}: $\text{Bernoulli}(\exp(-x))$ for general $x \geq 0$, implemented by repeatedly invoking \texttt{\seqsplit{sample\_bernoulli\_exp1}} on the integer and fractional parts of $x$

\item \texttt{sample\_bernoulli\_exp1}: $\text{Bernoulli}(\exp(-x))$ restricted to $x \in [0,1]$, based on an alternating Taylor series

\item \texttt{sample\_bernoulli\_rational}: $\text{Bernoulli}(p)$ for rational $p = a/b$

\item \texttt{sample\_uniform\_ubig\_below}: rejection sampler for uniform big integer on $\{0,\dots,\texttt{upper}-1\}$ based on repeated calls to the underlying byte-level RNG.
\end{itemize}

\section{Diagnostic Methodology}
\label{sec:diagnostic}

To localize the source of the artifacts, we adopted a top-down diagnostic approach. We implemented a traced version of OpenDP's CKS20 sampler that records representative outputs from every primitive in the call chain for each discrete Laplace sample generated.

Specifically, a single call to the traced \texttt{\seqsplit{sample\_discrete\_laplace}} returns both the final discrete Laplace sample and a diagnostic row containing: the first uniform value drawn; the first value from \texttt{\seqsplit{sample\_uniform\_ubig\_below}}; the first outcomes of \texttt{\seqsplit{sample\_bernoulli\_standard}}, \texttt{sample\_bernoulli\_rational}, \texttt{\seqsplit{sample\_bernoulli\_exp1}}, and \texttt{\seqsplit{sample\_bernoulli\_exp}}; the first magnitudes from \texttt{\seqsplit{sample\_geometric\_exp\_slow}} and \texttt{\seqsplit{sample\_geometric\_exp\_fast}}; and the final discrete Laplace sample.

By iterating $N = 10^6$ times, we saved these cascades of samples, all generated from the same entropy source. We then performed goodness-of-fit tests ($\chi^2$ and $z$-tests) on each column against the corresponding theoretical distribution. Because all columns derive from the same random bytes, any deviation must originate from one of the deterministic transformations. The key insight is that as soon as one primitive distorts the distribution, the discrepancy becomes visible in its column and propagates to all downstream columns, allowing us to pinpoint the earliest faulty step in the sampling pipeline.

\section{Layer-by-Layer Analysis}
\label{sec:analysis}

\subsection{Uniform Sampling}

We began by verifying the uniformity of the bit-stream and the rejection sampler function \texttt{\seqsplit{sample\_uniform\_ubig\_below}}. Results are presented in Figure \ref{fig:uniformity_bad}. Both distributions yield $\chi^2_{\text{red}} \approx 1$ and $p$-values of $p = 0.433$ (uniform bit sampling) and $p = 0.654$ (thresholded uniform sampler) when tested against the theoretical uniform model. These results confirm that the entropy source and uniform integer sampling layer operate correctly.

\subsection{Bernoulli Samplers}

The next layer consists of the Bernoulli samplers: \texttt{\seqsplit{sample\_standard\_bernoulli}}, \texttt{\seqsplit{sample\_bernoulli\_exp}}, \texttt{\seqsplit{sample\_bernoulli\_exp1}}, and \texttt{\seqsplit{sample\_bernoulli\_rational}}. The standard Bernoulli is called only by the Laplace sampler to determine the sign and draws directly from the uniform entropy source. The other three call each other sequentially in a nested environment, with only the rational sampler accessing the entropy source directly.

To test conformity, we performed both a $\chi^2$ goodness-of-fit test and a $z$-test on the empirical success probability $\hat{p} = k_1 / n$ against the theoretical value $p_{\text{th}}$, using the normal approximation to the binomial:
\[
z = \frac{\hat{p} - p_{\text{th}}}{\sqrt{p_{\text{th}}(1 - p_{\text{th}})/n}}
\]
The $z$-value represents the number of standard deviations the observed probability deviates from the expected value; values within $|z| < 3$ indicate agreement with the theoretical model at 99\% confidence.

The standard Bernoulli was tested against $p = 0.5$. The rational Bernoulli was tested against $p = 1/b = 0.1$ (matching the Laplace scale). Both exponential functions were tested against
\[
p = \mathbb{E}[e^{-U/b}] = \frac{1}{b} \sum_{u=0}^{b-1} e^{-u/b} = \frac{1}{b} \cdot \frac{1 - e^{-1}}{1 - e^{-1/b}},
\]
where $U$ is uniformly sampled from $\{0, \ldots, b-1\}$.

The \texttt{\seqsplit{sample\_standard\_bernoulli}} and \texttt{\seqsplit{sample\_bernoulli\_rational}} functions pass both tests. The standard Bernoulli returns $z = 0.58$ with $p = 0.56$, and the rational Bernoulli returns $z = 0.66$ with $p = 0.51$. In both cases, the $z$-value sits within $1\sigma$ of the expected value, with correspondingly high $p$-values (Figure \ref{fig:bernoulli_bad}).

The exponential functions, however, fail the tests dramatically. The \texttt{\seqsplit{bernoulli\_exp1}} function returns $z = 11.30$ with $p \approx 0$ and $\chi^2 = 127.7$. The \texttt{\seqsplit{bernoulli\_exp}} function shows identical statistics, which is expected given the calling hierarchy. Since function calls proceed from \texttt{\seqsplit{bernoulli\_exp}} to \texttt{\seqsplit{bernoulli\_exp1}} to \texttt{\seqsplit{bernoulli\_rational}}, the failing function is most likely \texttt{\seqsplit{bernoulli\_exp1}}, with the error propagating upward to \texttt{\seqsplit{bernoulli\_exp}} (Figure \ref{fig:bernoulli_exps_bad}).

\subsection{Geometric Samplers}

Moving to the geometric samplers, we observe a clear divergence. The slow geometric distribution appears correct, while the fast geometric sampler exhibits artifacts similar to those in the final Laplacian output and fails statistical tests.

We performed both $z$-tests and $\chi^2$ tests against the geometric distribution
\[
P(K = k) = (1 - e^{-x})\, e^{-xk}, \quad k = 0, 1, 2, \ldots
\]
where $x = 1$ for \texttt{\seqsplit{sample\_geometric\_exp\_slow}} and $x = 1/b$ for \texttt{\seqsplit{sample\_geometric\_exp\_fast}}.

The slow geometric sampler appears correct despite depending on the faulty \texttt{\seqsplit{bernoulli\_exp}} because it is always called with parameter $x=1$, meaning it only uses a single Bernoulli with $p=e^{-1}$, implemented via \texttt{\seqsplit{bernoulli\_exp1(1)}}. 
As discussed in Section \ref{sec:discussion}, the bug in the underlying \texttt{dashu} library produces errors that depend on the specific values involved. For this particular input the resulting bias is negligible, and the function returns a correct result.

The fast geometric sampler, in contrast, repeatedly samples from an entire family of slightly biased coins with probabilities $e^{-u/b}$. This over-selects some samples while under-selecting others, distorting the entire geometric distribution and causing test failures (Figure \ref{fig:geometric_bad}).

\subsection{Final Laplace Distribution}

The generated discrete Laplace distribution (Figure \ref{fig:discrete_laplace_bad}) exhibits the artifacts we first observed (mean $z = -0.68$ with $p = 0.49$; $\chi^2 = 2.1 \times 10^3$ with $p = 0$). While the mean is well-centered, the distribution fails goodness-of-fit testing due to systematic deviations. The top-down analysis has successfully traced the problem from its visible manifestation in the final output back to its origin in the \texttt{\seqsplit{bernoulli\_exp1}} function. In the following Section, we propose an alternative implementation of this sampler that eliminates the artifacts while leaving the rest of the CKS20 construction untouched.

\begin{algorithm}[t]
\caption{Sample $\text{Bernoulli}(\exp(-x))$ via Taylor Series}
\label{alg:bernoulli_exp}
\KwIn{$x \in [0,1]$}
\KwOut{True with probability $\exp(-x)$}
\If{$x = 0$}{\Return{True}}
Draw $U \sim \text{Uniform}(\{0, \ldots, 2^{128}-1\})$; $u \gets U / 2^{128}$

$S_{\text{prev}} \gets 1$; $S_{\text{curr}} \gets 1 - x$; $k \gets 1$; $t \gets x$

\While{True}{
$L \gets \min(S_{\text{prev}}, S_{\text{curr}})$; $H \gets \max(S_{\text{prev}}, S_{\text{curr}})$

\If{$u < L$}{\Return{True}}
\If{$u \geq H$}{\Return{False}}
$k \gets k + 1$; $t \gets t \cdot x / k$; $S_{\text{prev}} \gets S_{\text{curr}}$

\eIf{$k$ is even}{$S_{\text{curr}} \gets S_{\text{curr}} + t$}{$S_{\text{curr}} \gets S_{\text{curr}} - t$}
}
\end{algorithm}

\begin{figure}[t]
\centering
\includegraphics[width=0.7\columnwidth]{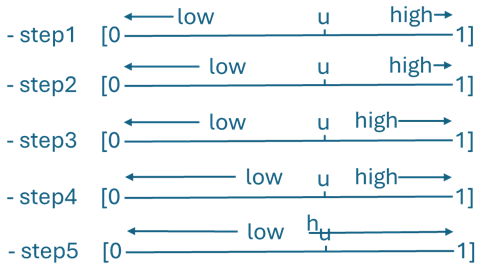}
\caption{Iterative refinement of the Taylor series bounds. At each step, the interval $[L, H]$ tightens around the true value. The algorithm terminates when $u$ falls outside the current bracket, returning True if $u < L$ or False if $u \geq H$.}
\label{fig:algorithm}
\end{figure}

\section{Alternative Sampling Algorithm}
\label{sec:proposed}

Our alternative implementation is based on the proof of Proposition 33 in CKS20 \cite{discrete_gauss_opendp} and on an alternating series Bernoulli sampler described in \cite{occil2025bernoulli}. The algorithm samples from a $\text{Bernoulli}(\exp(-x))$ distribution by iteratively refining the partial sums of the Taylor series expansion until a decision can be made.

The first step samples a 128-bit uniformly random integer $U \in \{0, \ldots, 2^{128}-1\}$ and interprets it as the rational $u = U / 2^{128}$. In our implementation, this fraction is stored as a pair of arbitrary-precision integers (numerator and denominator) using Rust's \texttt{dashu::RBig} type \cite{dashu}. 

The algorithm exploits the Taylor series expansion:
\[
e^{-x} = \sum_{k=0}^{\infty} \frac{(-x)^k}{k!} = 1 - x + \frac{x^2}{2!} - \frac{x^3}{3!} + \cdots
\]
Because this is an alternating series with decreasing terms, consecutive partial sums $S_{k-1}$ and $S_k$ always bracket the true value $e^{-x}$: one lies above and the other below.

The algorithm initializes $S_{\text{prev}} = 1$ and $S_{\text{curr}} = 1 - x$, corresponding to the first two partial sums. At each iteration, it computes bounds $L = \min(S_{\text{prev}}, S_{\text{curr}})$ and $H = \max(S_{\text{prev}}, S_{\text{curr}})$, then compares the uniform sample $u$ against these thresholds. If $u < L$, we know with certainty that $u < e^{-x}$, so the algorithm returns True. If $u \geq H$, we know $u \geq e^{-x}$, so it returns False. Otherwise, $u$ falls within the undecided region, and the algorithm computes the next term until convergence. Figure \ref{fig:algorithm} illustrates this process. At each step, the bounds converge toward $e^{-x}$ until $u$ falls outside the shrinking interval, at which point a decision is made.

\section{Validation Results}
\label{sec:results}

After replacing the original \texttt{\seqsplit{bernoulli\_exp1}} function with our Taylor series implementation, we repeated the diagnostic pipeline to verify that the artifacts had been resolved. The results confirm that all samples now correctly fit their theoretical distributions.

\begin{figure*}[t]
\centering
\begin{subfigure}[b]{0.4\textwidth}
\centering
\includegraphics[width=\textwidth]{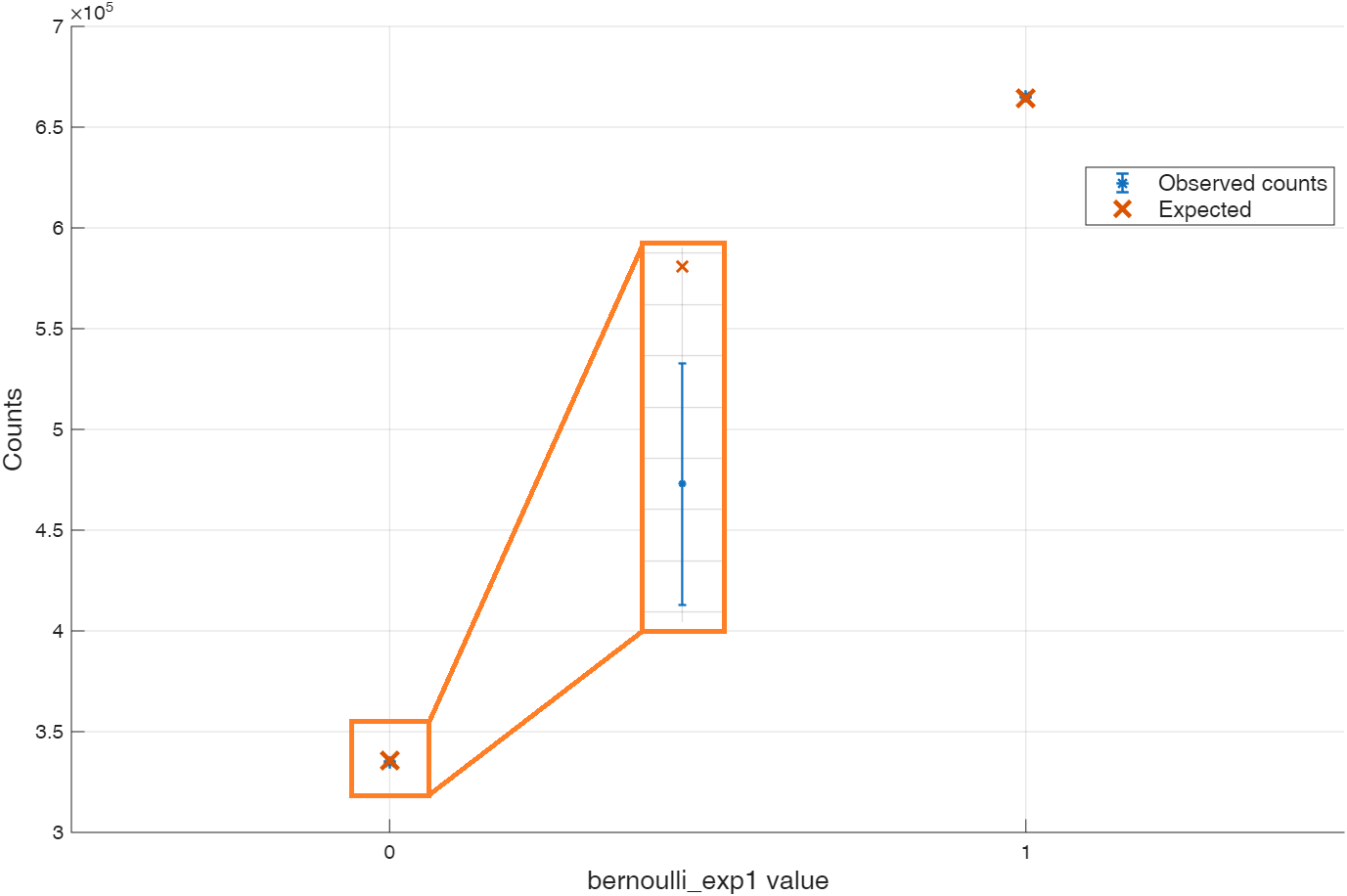}
\caption{\texttt{bernoulli\_exp1}: $z = 1.80$, $p = 0.073$.}
\label{fig:bernoulli_exp1_good}
\end{subfigure}
\hfill
\begin{subfigure}[b]{0.4\textwidth}
\centering
\includegraphics[width=\textwidth]{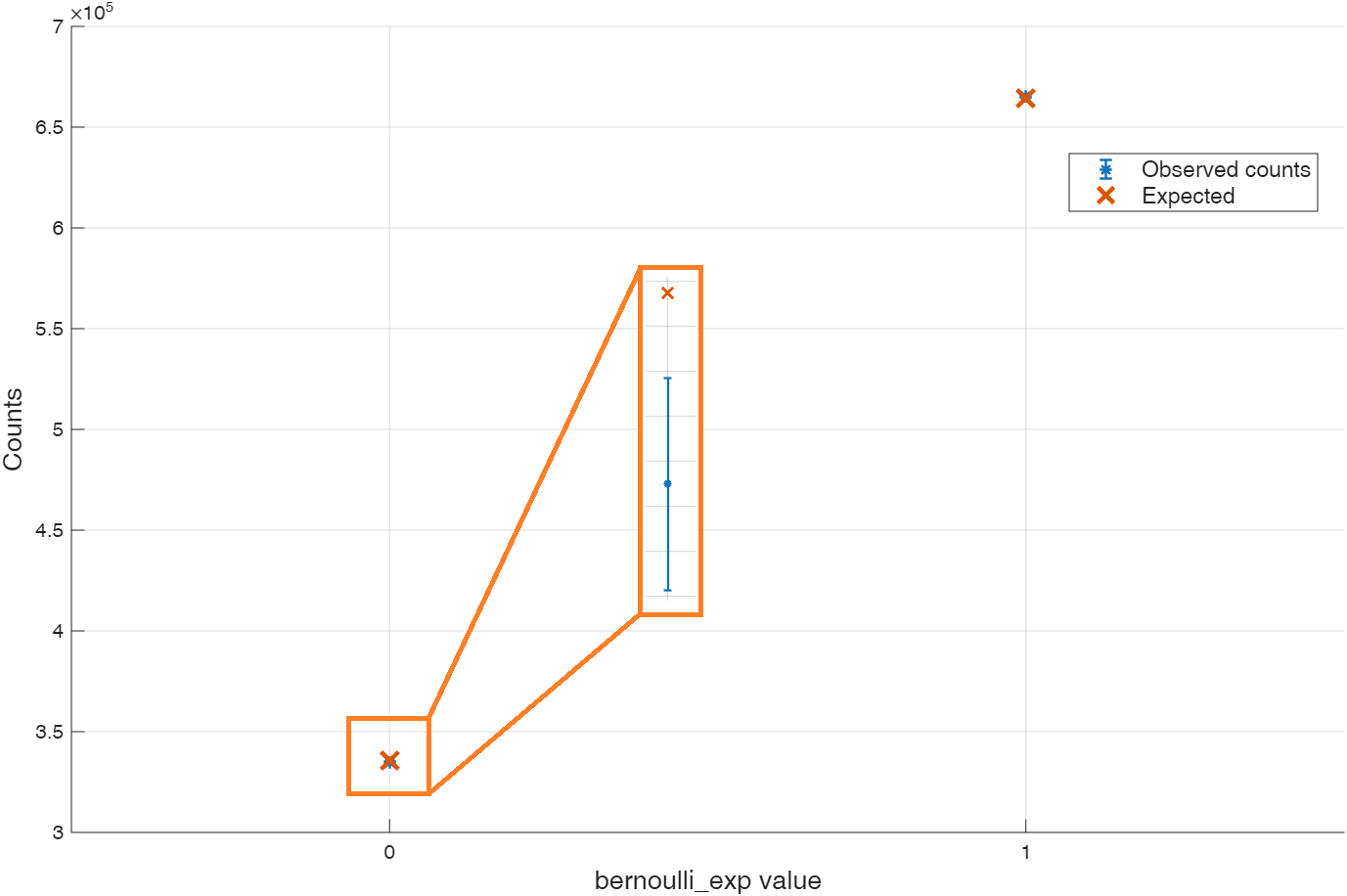}
\caption{\texttt{bernoulli\_exp}: $z = 1.80$, $p = 0.073$.}
\label{fig:bernoulli_exp_good}
\end{subfigure}
\caption{Exponential Bernoulli samplers with the new implementation. Left: \texttt{bernoulli\_exp} ($\hat{p} = 0.6651$, $p_{\text{th}} = 0.6643$, $\chi^2 = 3.22$, $p = 0.0726$). Right: \texttt{bernoulli\_exp1} ($\hat{p} = 0.6651$, $p_{\text{th}} = 0.6643$, $\chi^2 = 3.22$, $p = 0.0726$). Both now pass goodness-of-fit tests with $\chi^2_{\text{red}} \approx 3$ and $p > 0.07$, demonstrating correct sampling behavior. The insets on the plots show a zoomed region around the expected value, with error bars representing $1\sigma$ confidence intervals, showing how the observed values are within $3\sigma$ from the expected values.}
\label{fig:bernoulli_exps_good}

\begin{subfigure}[b]{0.32\textwidth}
\centering
\includegraphics[width=\textwidth]{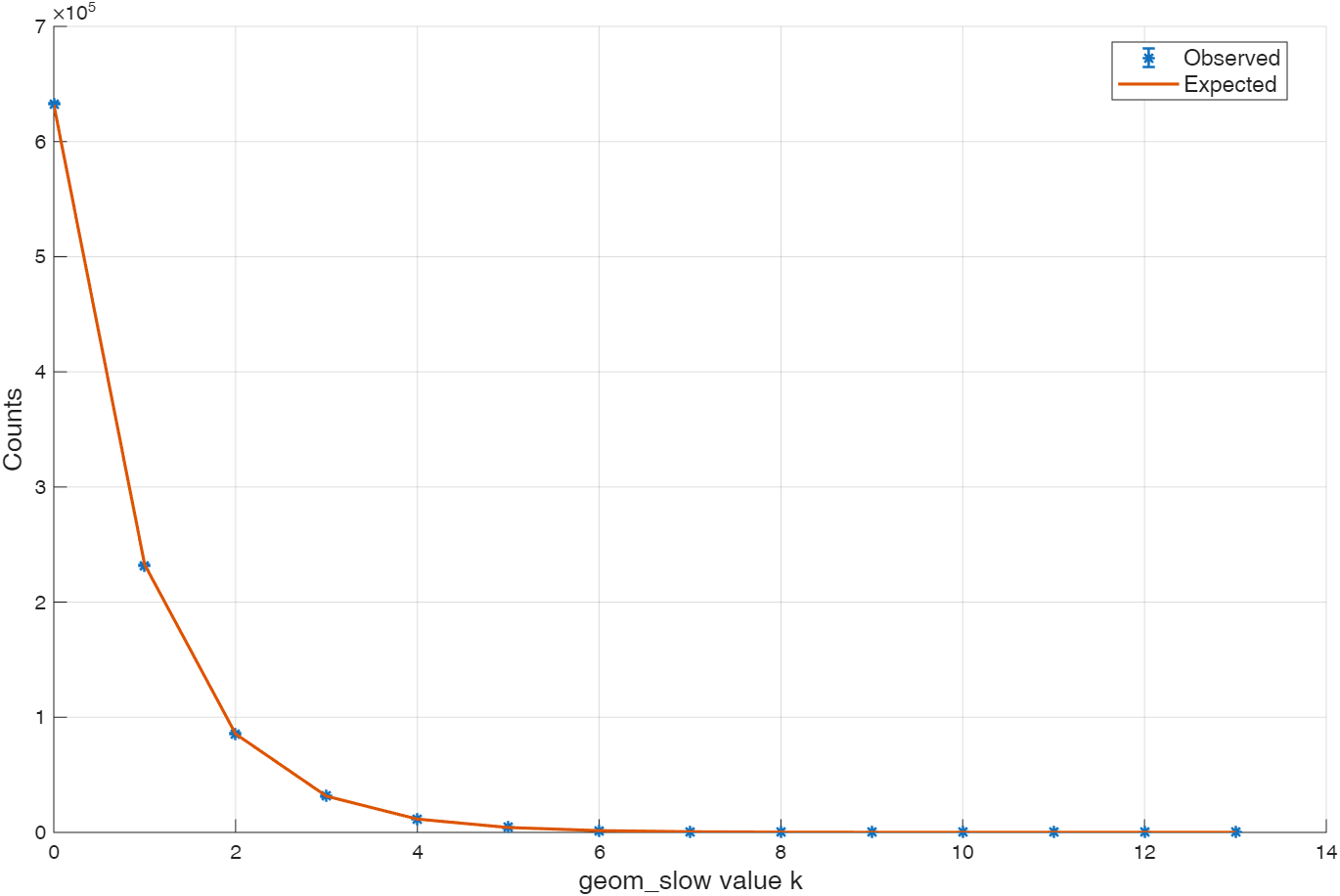}
\caption{Geometric slow: $\chi^2 = 9.74$, $p = 0.715$. ~~~~~~~~~~~~~~~~~~~~}
\label{fig:geometric_slow_good}
\end{subfigure}
\hfill
\begin{subfigure}[b]{0.32\textwidth}
\centering
\includegraphics[width=\textwidth]{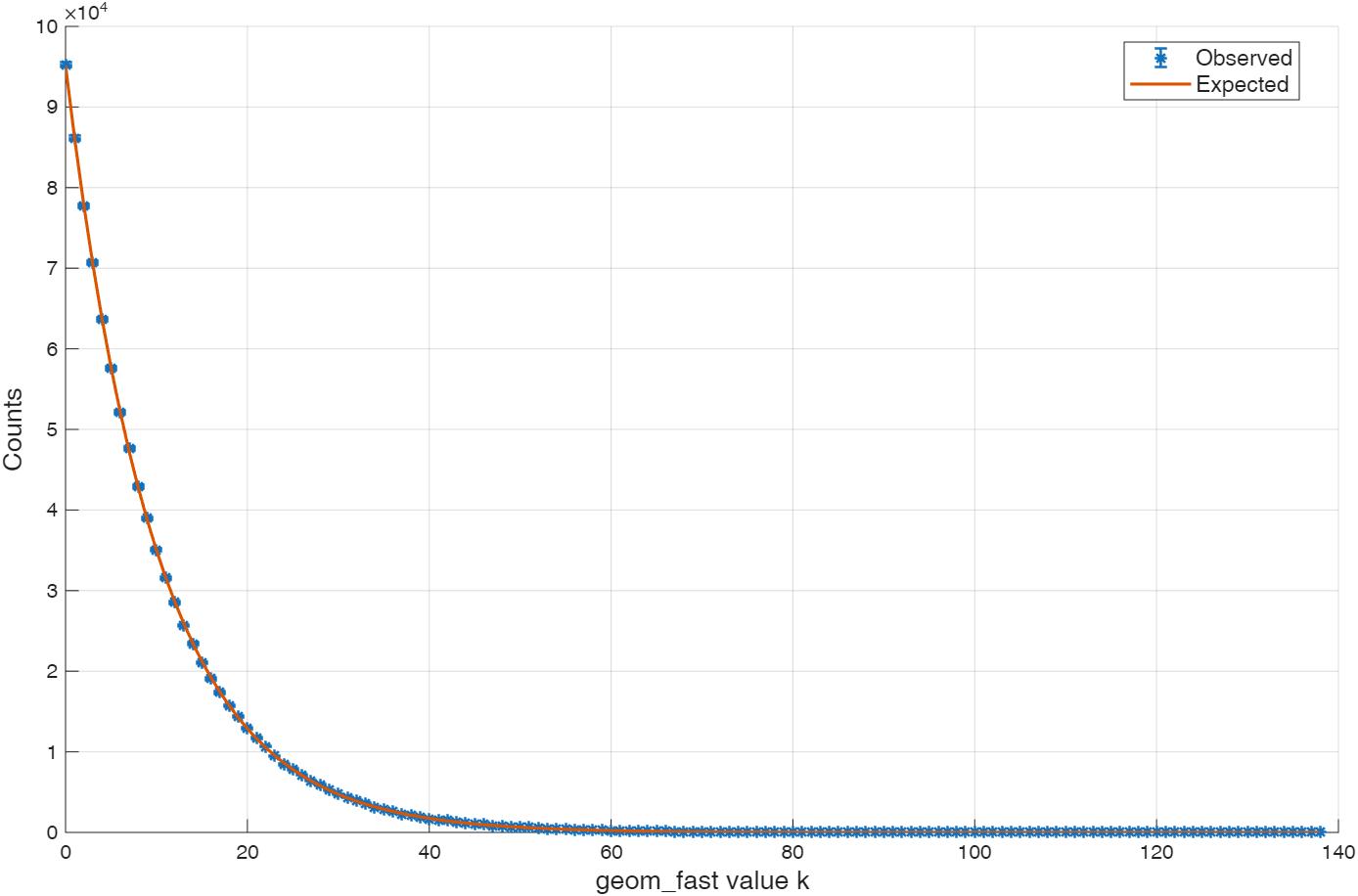}
\caption{Geometric fast: $\chi^2 = 145.5$, $p = 0.315$. ~~~~~~~~~~~~~~~}
\label{fig:geometric_fast_good}
\end{subfigure}
\hfill
\begin{subfigure}[b]{0.32\textwidth}
\centering
\includegraphics[width=\textwidth]{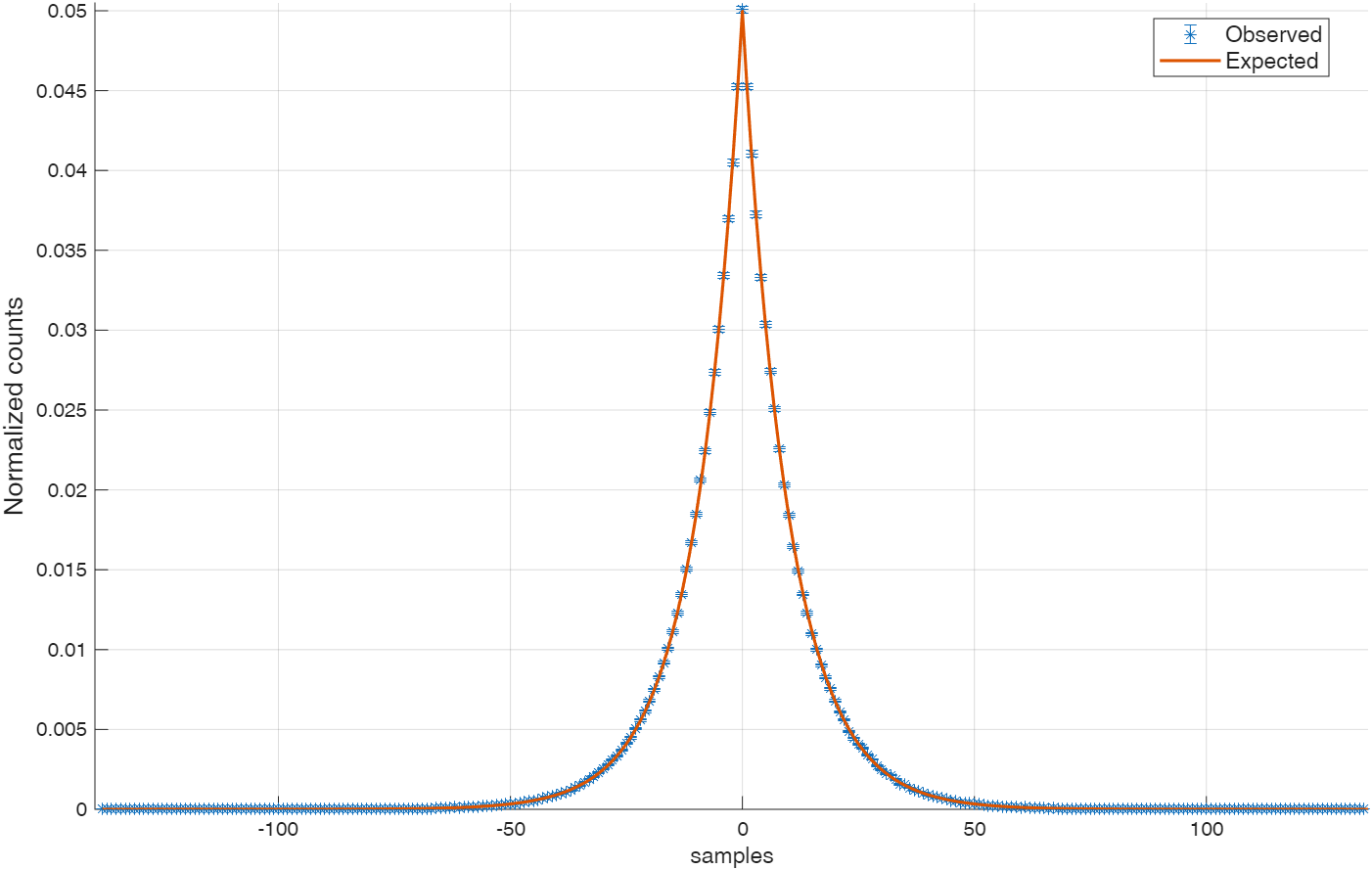}
\caption{Discrete Laplace: $z = -1.22$, $p = 0.22$; $\chi^2 = 2.8 \times 10^2$, $p = 0.5$.}
\label{fig:discrete_laplace_good}
\end{subfigure}

\caption{Geometric and Discrete Laplace sampler distributions with the corrected implementation. (a) Slow implementation (mean $z = -0.38$, $p = 0.71$; $\chi^2 = 9.7$, $p = 0.72$). (b) Fast implementation (mean $z = -0.18$, $p = 0.86$; $\chi^2 = 1.5 \times 10^2$, $p = 0.31$). (c) Discrete Laplace distribution (mean $z = -1.22$, $p = 0.22$; $\chi^2 = 2.8 \times 10^2$, $p = 0.5$). The periodic artifacts visible in Figure \ref{fig:discrete_laplace_bad} have been eliminated. All samplers pass goodness-of-fit tests.}
\label{fig:geometric_good}

\vspace{-0.5cm}
\end{figure*}

\subsection{Bernoulli Samplers}

Figure \ref{fig:bernoulli_exps_good} shows that both \texttt{bernoulli\_exp1} and \texttt{bernoulli\_exp} now produce samples consistent with the theoretical model. From $n = 10^6$ samples, we observe an empirical success probability of $\hat{p} = 0.6651$ against a theoretical value of $p_{\text{th}} = 0.6643$. The $z$-test yields $z = 1.80$ with $p = 0.073$, and the $\chi^2$ test gives $\chi^2 = 3.22$ with $p = 0.073$. While the $p$-values are somewhat low, the $z$-score sits within $2\sigma$, indicating acceptable statistical fluctuation rather than systematic bias. This stands in stark contrast to the original implementation, which produced $z \approx 11$ and $p \approx 0$. Results were confirmed through independent experiments with different parameter choices.

\subsection{Geometric Samplers}

The improvement propagates through the sampling hierarchy. Figure \ref{fig:geometric_good} shows the geometric samplers, both of which now pass goodness-of-fit tests. The slow geometric sampler continues to perform well ($\chi^2 = 9.74$ with 13 degrees of freedom, $p = 0.715$). The fast geometric sampler now adheres to the theoretical distribution ($\chi^2 = 145.5$ with 138 degrees of freedom, $p = 0.315$), with empirical moments closely matching theoretical values: $\hat{\mu} = 9.507$ against $\mu_{\text{th}} = 9.508$, and $\hat{\sigma}^2 = 99.94$ against $\sigma^2_{\text{th}} = 99.92$. The periodic artifacts are no longer visible.

\subsection{Discrete Laplace Distribution}

Figure \ref{fig:discrete_laplace_good} shows the discrete Laplace distribution produced by the corrected sampler. The characteristic periodic artifacts (compare with Figure \ref{fig:discrete_laplace_bad}) have been completely eliminated. The distribution passes the $\chi^2$ goodness-of-fit test with $\chi^2 = 275.4$ on 276 degrees of freedom ($p = 0.499$), indicating excellent agreement with the theoretical model. The empirical variance $\hat{\sigma}^2 = 199.90$ matches the theoretical value $\sigma^2_{\text{th}} = 199.83$, and the sample mean $\hat{\mu} = -0.017$ is statistically indistinguishable from zero ($z = -1.22$, $p = 0.224$).

These results confirm that the artifacts originated from the \texttt{bernoulli\_exp1} implementation and that our proposed fix seemingly resolves the issue. The corrected sampler produces discrete Laplace samples that are statistically indistinguishable from the theoretical distribution at the tested precision level.

\begin{figure}[htp]
  \centering
  \subfloat[Original]{%
    \includegraphics[width=\columnwidth]{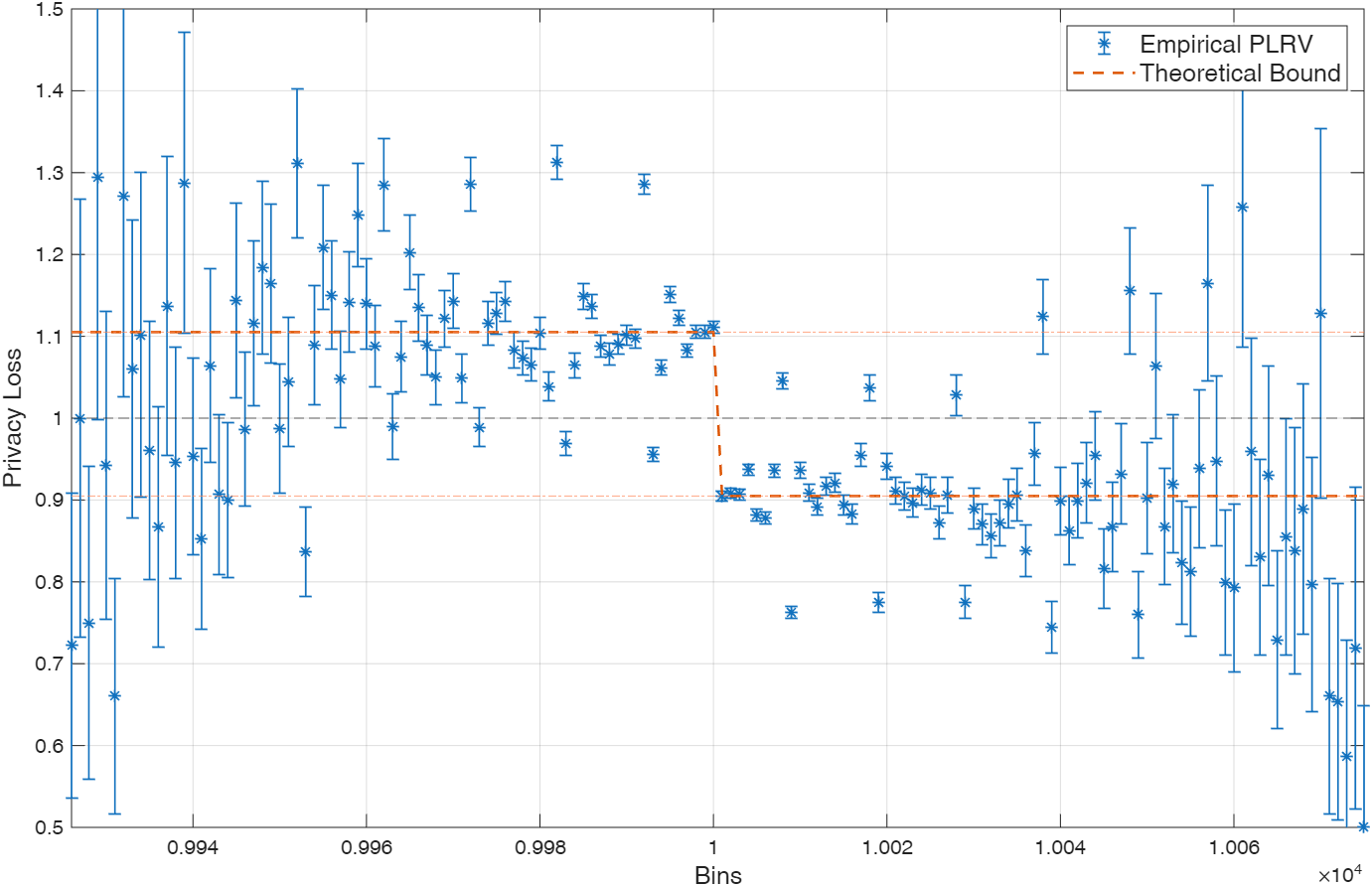}%
    \label{fig:plrv-original}}
  \hfill
  \subfloat[Taylor series]{%
    \includegraphics[width=\columnwidth]{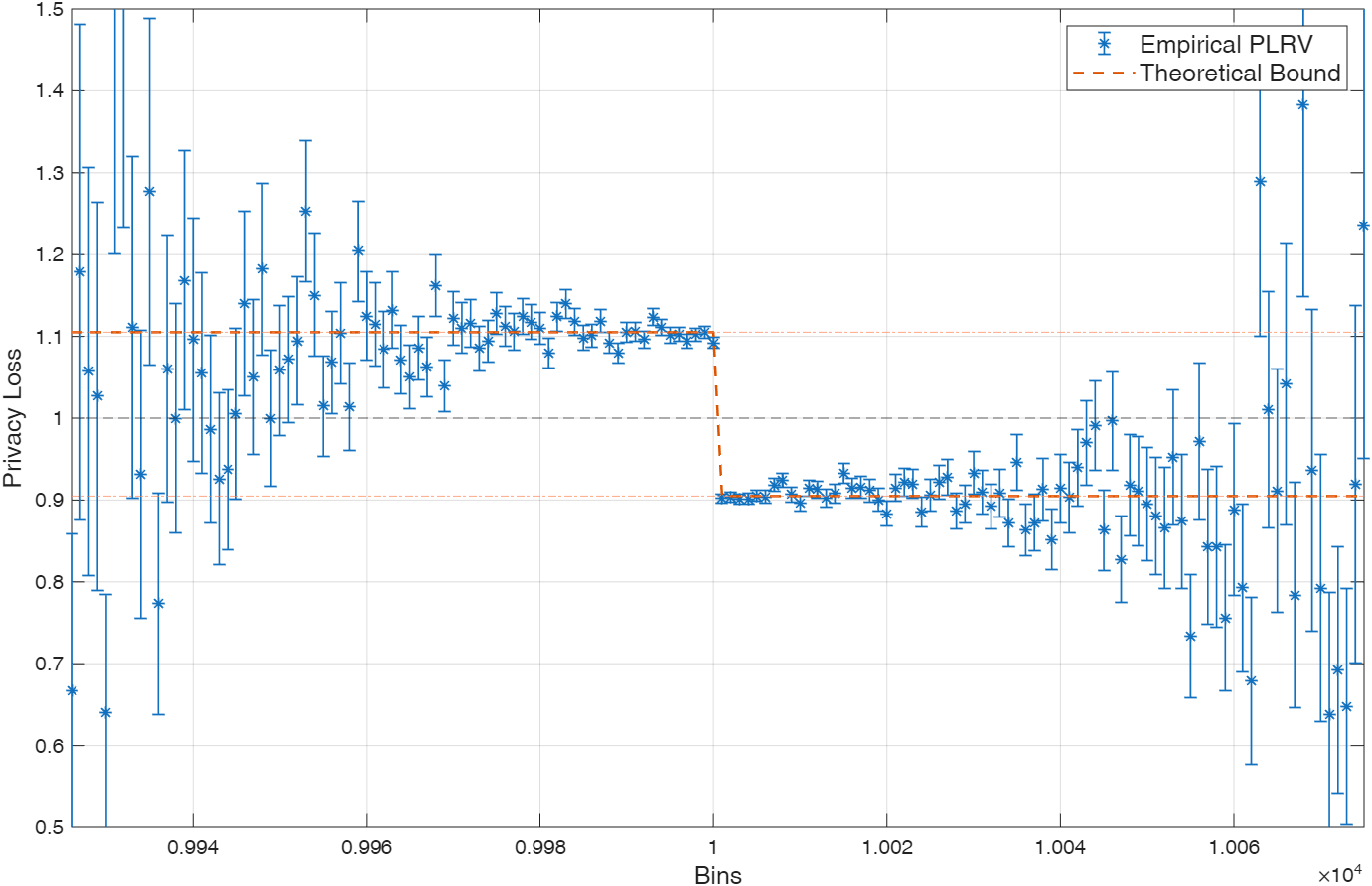}%
    \label{fig:plrv-taylor}}
  \hfill
  \subfloat[OpenDP v0.14.2]{%
    \includegraphics[width=\columnwidth]{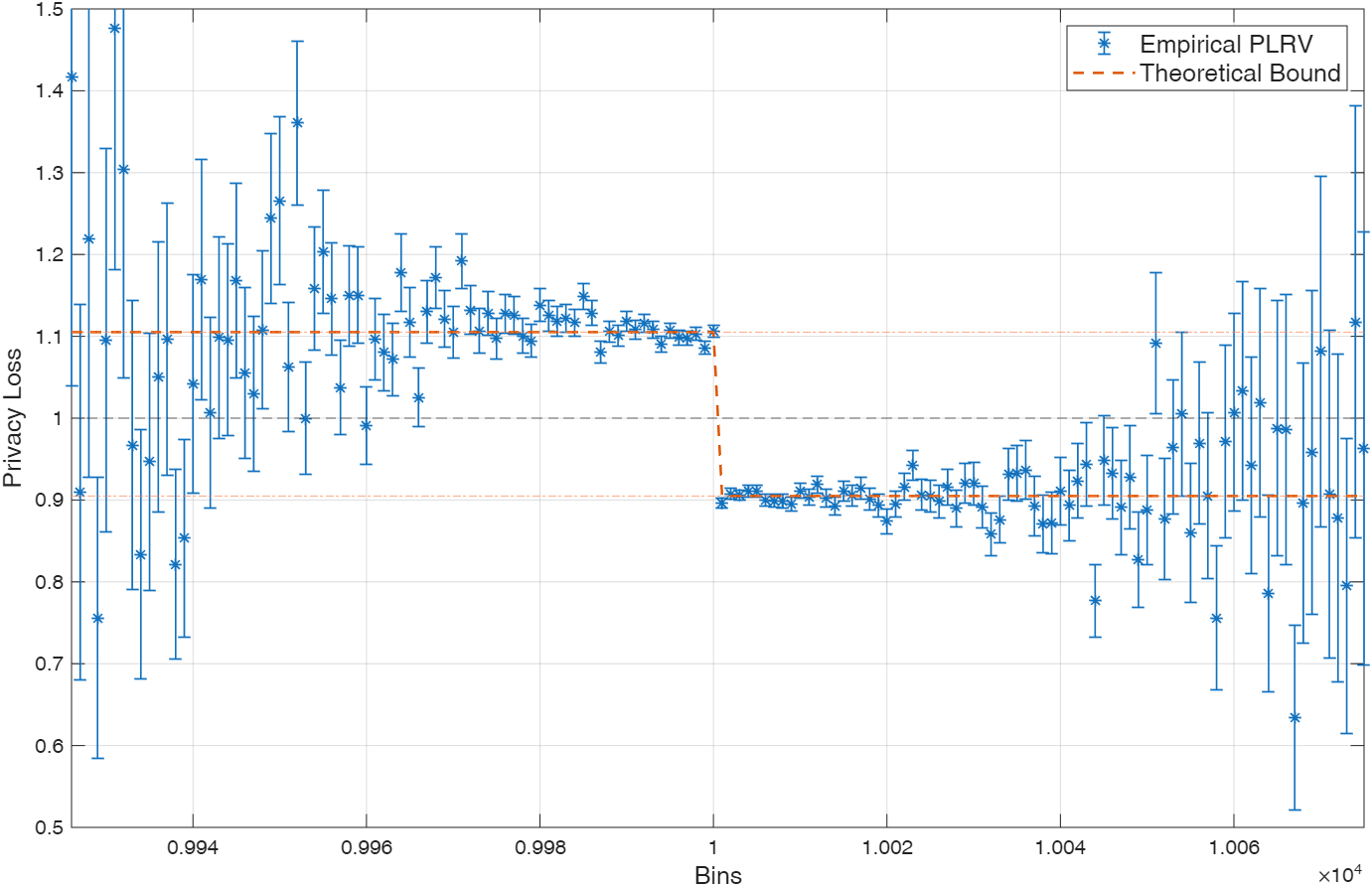}%
    \label{fig:plrv-fixed}}
  \caption{Empirical PLRV from the output distributions of two neighboring datasets ($\varepsilon = 0.1$, $N = 10^6$, bin size $= 1$). Dashed orange lines highlight the theoretical bounds $e^{\pm\varepsilon}$. Error bars are $1\sigma$ confidence intervals. The faulty implementation (a) shows points scattered beyond $e^{\pm\varepsilon}$, while both corrections (b, c) remain within bounds at $3\sigma$ confidence level.}
  \label{fig:plrv-comparison}

\end{figure}

\section{Discussion}
\label{sec:discussion}

Following disclosure with the OpenDP development team, the root cause of the artifacts 
was identified as an issue in the \texttt{dashu} library's \texttt{RBig / UBig} operator 
\cite{dashu_issue57}. 

Each coin flip requires evaluating $\text{Bernoulli}(x/k)$, which internally calls \texttt{\seqsplit{sample\_bernoulli\_rational}} with $x/k$ as a rational argument. This division is delegated to the faulty operator. The published release of \texttt{dashu} (v0.4, January 2024) contains a bug in this operator: given $x = a/b$ and divisor $k$, the implementation computes the denominator as $(b/g) \cdot k$ instead of the correct $b \cdot (k/g)$, where $g = \gcd(a, k)$ \cite{dashu_issue57}. The issue has been patched in the \texttt{dashu} master branch \cite{dashu_fix} but was never included in an official release.

As a result, each call to \texttt{\seqsplit{sample\_bernoulli\_rational}} within \texttt{\seqsplit{sample\_bernoulli\_exp1}} receives a subtly incorrect probability. These errors depend on the value of $k$, producing a bias in the coin flip outcomes. When accumulated in the loop, they are then propagated through the geometric and discrete Laplace constructions, resulting in the artifacts.

The OpenDP development team bypassed the call to the division by replacing it with a local implementation that computes $b \cdot (k/g)$ directly, preserving the full implementation of CKS20. The fix was integrated through pull request \#2639, and pushed in the official release v0.14.2 \cite{opendp_pr_2639}.

We independently repeated the diagnostic pipeline of Section \ref{sec:analysis} on the patched implementation, and all samplers in the chain pass the goodness-of-fit tests, with no visible artifacts in the final discrete Laplace distribution. We therefore continue this section with a brief performance benchmark of the three implementations and a remark on the practical privacy implications of the original faulty implementation.

\subsection{Performance Benchmarking}

While statistical correctness makes the OpenDP fix and the alternative Taylor-series sampler equivalent, their computational complexity differs. The original and the patched implementations follow the same CKS20 structure and differ only in where the rational division $x/k$ is carried out (delegated to \texttt{dashu} in the former, performed locally in the latter) so we expect their computational cost to be essentially the same. The Taylor sampler, on the other hand, replaces the rejection scheme with a looping alternating-series construction.

We measured the mean time per call to \texttt{\seqsplit{sample\_discrete\_laplace}} over a run of $10^6$ samples each. The original sampler took $14.68 \pm 0.01 \mu\text{s}$, the Fixed sampler $15.08 \pm 0.01 \mu\text{s}$, and the Taylor sampler $17.45 \pm 0.01 \mu\text{s}$. 

The original and fixed costs sit within $2.7\%$ of each other, indicating that the local implementation of the division does not change the complexity of the algorithm, as expected. The Taylor sampler is $15.7\%$ slower than the fixed implementation, but remains informative: two structurally different fixes output the same empirical behavior, which suggests that the problem was correctly localized in \texttt{\seqsplit{sample\_bernoulli\_exp1}}.

\subsection{Implications on the Privacy Guarantee}

A natural question is whether the artifacts described translate into a measurable degradation of the theoretical privacy guarantee. The privacy guarantees of CKS20 are established for noise drawn from the exact discrete Laplace distribution\cite{discrete_gauss_opendp}. Once the underlying $\mathrm{Bernoulli}(\exp(-\gamma))$ primitive is biased, those bounds are no longer guaranteed to apply.

To examine the consequences in practice, we employed the operational definition of Differential Privacy \cite{dp_book}, and reconstructed the empirical Privacy Loss Random Variable (PLRV), a precise tool to measure the privacy guarantees of a differentially private mechanism\cite{desfontainesblog20200306}.

For a randomized mechanism $\mathcal{M}$ and a pair of neighboring datasets $D_1, D_2$, the privacy loss at output $y$ is defined as 
\[
\mathcal{L}_{\mathcal{M}}^{D_1, D_2}(y) =
\ln \left( \frac{P(\mathcal{M}(D_1) = y)}{P(\mathcal{M}(D_2) = y)} \right),
\]
and a mechanism is $\varepsilon$-DP if $|\mathcal{L}_{\mathcal{M}}^{D_1, D_2}(y)| \leq \varepsilon$. The PLRV thus provides a direct measurement of the privacy guarantee for each output: any bin whose value exceeds the $\pm\varepsilon$ band corresponds to an output for which the privacy guarantee is not held.

We executed $N = 10^6$ counting queries on two neighboring datasets $D_1$ and $D_2$ of sizes $10{,}000$ and $10{,}001$ respectively. The counting query returns the number of entries in the database and has unit sensitivity ($\Delta s  = 1$), since adding or removing a single entry changes the result by at most one. We set $\varepsilon = 0.1$, and estimated the PLRV as the bin-wise ratio of the resulting output  histograms.
We work with the ratio rather than its logarithm for computational simplicity, with the
corresponding bound becoming $e^{\pm\varepsilon}$ instead of $\pm\varepsilon$. Under correct $\varepsilon$-DP the empirical PLRV must lie, up to statistical fluctuations, within $e^{\pm\varepsilon}$ across all bins.

Figure \ref{fig:plrv-comparison} compares the empirical PLRV obtained with the three implementations. For the original (faulty) sampler (Figure \ref{fig:plrv-original}), the fraction of bins whose PLRV deviates from the theoretical bound by more than $3\sigma$ is well above the $\sim 0.3\%$ rate expected under Gaussian fluctuations, suggesting that the violations are systematic rather than random. Moreover, the bins that visibly deviate from the model can be traced back to the periodic artifacts identified in Figure \ref{fig:discrete_laplace_bad}: the same outputs that were over and under-sampled by the faulty Laplace distribution are those for which the $\varepsilon$-DP bound is violated. In other words, the artifacts visible in the noise distribution propagate to the PLRV and translate into visible violations of the $\varepsilon$-DP bound.

Both the Taylor-series sampler (Figure \ref{fig:plrv-taylor}) and the official fix (Figure \ref{fig:plrv-fixed}) produce PLRV values that remain within the theoretical bounds at $3\sigma$ confidence across the full histogram, with exceedances compatible with the expected statistical fluctuations.

We present this as a conjecture based on the empirical evidence above, not as a rigorous PLRV analysis. Turning the qualitative observation that the bound is violated into a quantitative statement about the effective privacy violation  would require a dedicated privacy-loss accounting analysis, which we do not attempt here.

\section{Conclusion}

We have identified and resolved systematic artifacts in OpenDP's discrete Laplace sampler through a systematic diagnostic methodology that isolated the root cause to numerical precision issues in the \texttt{bernoulli\_exp1} primitive. Our alternative implementation eliminates these artifacts at the tested precision level, in agreement with the official OpenDP patch released after our disclosure. Empirical analysis of the privacy loss suggests the artifacts also violate the $e^{\pm\varepsilon}$ bound under the faulty sampler, while both corrections restore it.

More broadly, this work demonstrates the importance of rigorous statistical validation for differential privacy implementations. Even mathematically correct algorithms can produce unexpected behaviors when translated to finite-precision systems, and subtle implementation details can compromise privacy guarantees or degrade utility. Our diagnostic approach provides a template for isolating similar issues in hierarchical sampling implementations and emphasizes the need for comprehensive testing beyond unit tests of individual components. This analysis was shared with the OpenDP development team, leading to the identification of the bug in the \texttt{dashu} rational arithmetic library \cite{dashu_issue57}, and for subsequent integration of extended statistical testing throughout the OpenDP library in response to this disclosure \cite{opendp-issue-2638}.

\section*{Acknowledgements}
We would like to thank Michael Shoemate\footnote{\url{https://github.com/Shoeboxam}} of the OpenDP development team for his valuable feedback and contributions following this disclosure. 

This project was partially funded by the European Union’s ``Next Generation EU" Program under the Piano Nazionale di Ripresa e Resilienza (PNRR) through NQSTI Spoke8.

\bibliographystyle{unsrt}
\bibliography{bibliography}

\end{document}